\documentclass[summary]{ursirfi}
\usepackage{hyperref}
\usepackage{float}
\usepackage{comment}
\hypersetup{
  colorlinks   = true,    
  urlcolor     = blue,    
  linkcolor    = blue,    
  citecolor    = red      
}

\title{Nature and Evolution of UHF and L-band Radio Frequency Interference at the MeerKAT Radio Telescope}

\author{Isaac Sihlangu*\affref{ref1}, 
Nadeem Oozeer\affref{ref1}\affref{ref2}\affref{ref5},
 and Bruce A. Bassett\affref{ref2}\affref{ref3}\affref{ref4}}

\affiliation{%
  \aff{ref1}{South African Radio Astronomy Observatory, Cape Town, South Africa}
  \aff{ref2}{African Institute for Mathematical Sciences, Muizenberg, South Africa}
  \aff{ref3}{University of Cape Town, Department of Maths and Applied Maths, Cape Town, South Africa}
  \aff{ref4}{South African Astronomical Observatory, Cape Town, South Africa}
  \aff{ref5}{Department of Physics and Electronics, Rhodes University, P.O. Box 94, Makhanda 6140, South Africa}
}

\begin{document}

\maketitle

\begin{abstract}

Radio Frequency Interference (RFI) is unwanted noise that swamps the desired astronomical signal. Radio astronomers have always had to deal with RFI detection and excision around telescope sites, but little has been done to understand the full scope, nature and evolution of RFI in a unified way. We undertake this for the MeerKAT array using a probabilistic multidimensional framework approach focussing on UHF-band and L-band data. In the UHF-band, RFI is dominated by allocated Global System for Mobile (GSM) Communications, flight Distance Measuring Equipment (DME), and UHF-TV bands. The L-band suffers from known RFI sources such as DMEs, GSM, and the Global Positioning System (GPS) satellites. In the ``clean'' MeerKAT band, we noticed the RFI occupancy changing with time and direction for both the L-band and UHF band. For example, we saw a significant increase (300\% increase) in the fraction of L-band flagged data in November 2018 compared to June 2018. This increase seems to correlate with construction activity on site. In the UHF-band, we found that the early morning is least impacted by RFI and other outliers. We also found a dramatic decrease in DME RFI during the hard lockdown due to the COVID-19 pandemic. The work presented here allows us to characterise the evolution of RFI at the MeerKAT site. Any observatory can adopt it to understand the behaviour of RFI within their surroundings.

\end{abstract}

\section{Introduction}

Radio astronomy is a branch of astronomy that studies astronomical objects by detecting the faint radio electromagnetic (EM) waves that they emit. Radio observatories are facing the enormous challenges posed by the advancement of the telecommunication world because of radio emissions from man-made activities that are orders of magnitude brighter than that of astronomical sources \cite{Morello}.

Over recent years, there has been a significant advancement in radio astronomy instruments, resulting in an incredible increase in the sensitivity of the current generation of telescopes such as the Lofar \cite{van Haarlem MP}, MeerKAT \cite{Jonas}, and ASKAP \cite{Banyer} compared to older generation telescopes. The increased sensitivity is a blessing to astronomers because it means resolving faint sources, however, it is also a curse because it makes the radio telescopes more susceptible to Radio Frequency Interference (RFI). Therefore, it is paramount that radio observatories keep track of, and understand, their RFI environment and to monitor for any changes in RFI. 

Several mechanisms have been implemented to deal with the unwanted signals from human activities that interfere with signals of extraterrestrial origin \cite{van Driel}. Such mechanisms include, but are not limited to; government regulations to protect radio astronomy observatories \footnote{\url{https://www.dst.gov.za/images/pdfs/SARAS\%20regs\%20intention.pdf}} and spectrum management policies that prohibit some of the radio emission, put restrictions on allowed power levels of RFI sources and also facilitate access to
the alternative radio communication channels to the local people  \cite{Morello}.

Nevertheless, all the mechanisms described above do not entirely prevent the negative impact on astronomical observations from these artificial signals. Nearly all the data recorded onto disk is contaminated by signals from human activities. Extra processing is required to detect the presence of these man-made signals. Astronomers use flagging techniques to mitigate RFI before carrying out any science. The concept of removing RFI from observational data has been vigorously dealt with, and several automated ways including machine learning methods have been explored and developed \cite{Morello}. The MeerKAT telescope, adopted the classic AOFlagger \cite{Offringa et al.} algorithm for its RFI post-processing although it is also possible to use more sophisticated deep learning algorithms that offer significant advantages including being adapted to the specific observatories \cite{deep}.

To understand the nature and the evolution of the RFI culprits at the Karoo site, we previously developed a multidimensional statistical approach \cite{sihlangu} and applied it to  MeerKAT data. The MeerKAT Historical Probability of RFI (KATHPRFI) framework computes the RFI probability as a function of five dimensions: time, frequency, baseline, azimuth and elevation(T, F, B, Az, El). 

The product of the KATHPRFI framework can be used to monitor occupancy of the dedicated frequency bands and provide alerts if a certain critical threshold is exceeded. Furthermore, the results from the KATHPRFI can be utilised to provide feedback on the data pipeline and optimise the flagging strategies based on the statistics. Finally, the product offers a way to keep track of the RFI environment over time. 

The RFI environment around the MeerKAT is dominated by  groups of known RFI. In the UHF-band (580 - 1015 MHz) these are: Global System for Mobile Communication (GSM), Distance Measurement Equipment (DME) and UHF TV transmitters (U-TV). The MeerKAT L-band (856 - 1711 MHz) is affected by the GSM, DME, Global Positioning System (GPS) and Global Navigation Satellite System (GNSS) bands.

These RFI bands are sub-divided as follows. The UHF television (UHF TV) stations transmit around 470 - 854 MHz \footnote{\url{ https://www.gov.za/sites/default/files/gcis\_document\\/201409/315011272b.pdf}}. The GSM band is divided into two significant sub-bands, the GSM-900 and GSM-1800. GSM-900 uses 890 - 915 MHz to send information from the Mobile Station to the Base Transceiver Station (uplink) and 935 - 960 MHz for the other direction (downlink). GSM-1800, on the other hand, uses 1710 - 1785 MHz to send information from the Mobile Station to the Base Transceiver Station (uplink) and 1805 - 1880 MHz for the other direction (downlink). The downlink frequency range is outside the MeerKAT UHF-band and the L-band.

DME are signals from aircraft transmitted in the frequency range of 962 MHz - 1213 MHz. In this paper, we group the air-to-ground and ground-to-air DME transmission into one group, DME.

This paper will look into a general overview of the MeerKAT RFI in the UHF and L-band. We will also investigate the evolution of the RFI over time such that RFI prevention mechanisms can be adapted if required. In particular we examine the impact of lockdown during COVID-19. In Section \ref{uhf}, we start by giving a broad overview of the MeerKAT RFI in the UHF-band following the similar approach used in \cite{sihlangu}. In section \ref{evo} we look into the evolution of the MeerKAT RFI environment in the L-band.

\section{MeerKAT UHF-band RFI Environment Overview}
\label{uhf}

We searched the MeerKAT observation archive for all files with the description ``RFI scanning with calibrators''. These are short imaging observations (5min) of a sample of southern radio calibrators scattered around the sky. The advantages of such short observations are that the resulting datasets are relatively small, and the antennas cover a wide range of Azimuth and Elevation. We removed all results that were pointing observations since they do not contain imaging tracks required for the KATHPRFI. We further filtered off all the observations that did not include 4096 frequency channels. The search resulted in 81 files for the duration from the $4^{th}$ March 2020 to $4^{th}$ November 2020. The total time of all these UHF-imaging datasets that the analysis was performed is 108 hours. We also noticed short observation files between the 22 till 04 hours. This is common, since these time are mostly used for longer dedicated science observations. As a result we will have less information about the RFI environment around those times. On average, we had around 27 antennas in the observing run, with a minimum of 23 antennas and a maximum of 31 antennas. The mode number of antennas was 29, and 26\% of our observations were around night/early morning time.

\begin{figure}[H]
    \centering
    \includegraphics[width=90mm]{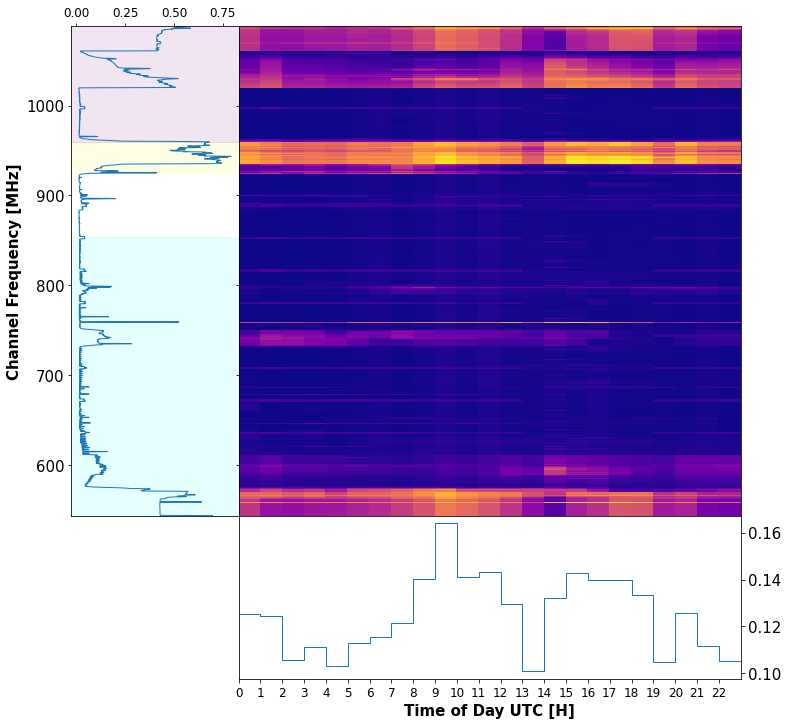}
    \caption{The distribution of the RFI probability (HH-pol)  as a function of frequency and Time of the day (in UTC). The colour code ranges from 0 to 1, with 0 corresponding to blue and 1 being yellow. The colours in the frequency sub-plot show regions of known RFI such as UHF TV (blue), GSM (yellow) and DME(pink) .}
    \label{time_freq}
\end{figure}

Given the three culprits of known RFI sources in the UHF-band, we investigated the distribution of RFI as a function of time of the day and frequency, and the results are shown in Figure \ref{time_freq}. The GSM, DME and UHF TV bands are shown in different colour bands.

From Figure \ref{time_freq}, the three central frequency regions are populated with outliers as expected. We notice a peak region around 544-583 MHz, which is the UHF TV. A narrow band outlier around 759 MHz is noted, that corresponds to the same band as the International Mobile Telecommunications  IMT700 BTX (Cellphone Downlink)\footnote{\url{https://www.nab.org.za/uploads/files/Radio\_Frequency\_Spectrum\_Assignment\_Plan\_Final\_C.pdf}}. 

Overall, we can confirm that these findings are consistent with the claim that the significant RFI  sources in the UHF-band at the  MeerKAT site are coming from the UHF TV, the GSM and the DME. The peak RFI is around 09:00-12:00 UTC (11:00-14:00 South African Standard Time - SAST) with a gap of an hour and peaks again between 16:00 to decrease after 21:00 SAST. A maximum variation of $6\%$ is observed between hours of the day in the RFI occupancy with an average of $13\%$. These results does not reveal a clear pattern between hours of the day and the fraction of data flagged. 

\begin{figure}[H]
    \centering
    \includegraphics[width=90mm]{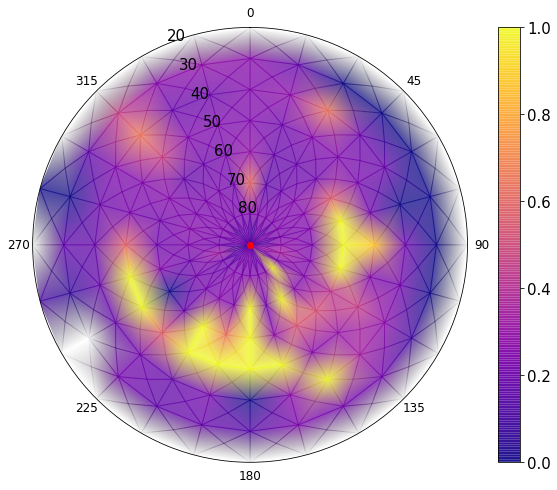}
    \caption{A polar plot of the RFI probability (VV-pol)  as a function of azimuth (radial direction) and elevation (theta direction). The white region at lower elevation is due to the lack of data availability in this range.}
    \label{polar}
\end{figure}

We investigated how the maximum amount of data flagged changes as a function of the telescope pointing in the UHF band. Figure \ref{polar} shows the results from the analysis. We found a hot spot region centred around Az = 90\textdegree bin for El between 50\textdegree - 60\textdegree \, bin. This can be associated with the cell phone towers at Carnarvon, which is not only the closest town to the MeerKAT site but is also situated around this bearing from site. In addition, we noticed couple other hot spots between Az 135\textdegree to Az 270\textdegree \, at higher elevations angles (50\textdegree to 80\textdegree). We can not ascertain the exact direction of these RFIs as they may be falling on any part of the lobes of the antenna pattern.

\section{MeerKAT RFI Evolution in the L-band}
\label{evo}

To study the MeerKAT RFI evolution in the L-band, we used the same dataset as in \cite{sihlangu}. This dataset corresponds to approximately 1500 hours of MeerKAT data ($\approx$200 TB) that was collected from May 2018 to January 2019. The results in \cite{sihlangu} showed that the three regions in the MeerKAT spectrum where RFI occurrence is the highest corresponds to the allocated GSM, DME, and GPS bands. Outliers from these sources are extremely harmful to science observations because of their 100\% duty cycle and band usage. Hence, recovering scientific quality observation can be almost hopeless. 

One always expect RFI to be lower at night as compared to daytime. However, in the MeerKAT L-band, we can not say so due to the sources of the RFI. From our time analysis, we found that there is not much difference (~ 3\%) in outlier occupancy as a function of the time of the day \cite{sihlangu}. 

\begin{figure}[H]
  \centering
  \includegraphics[width=80mm]{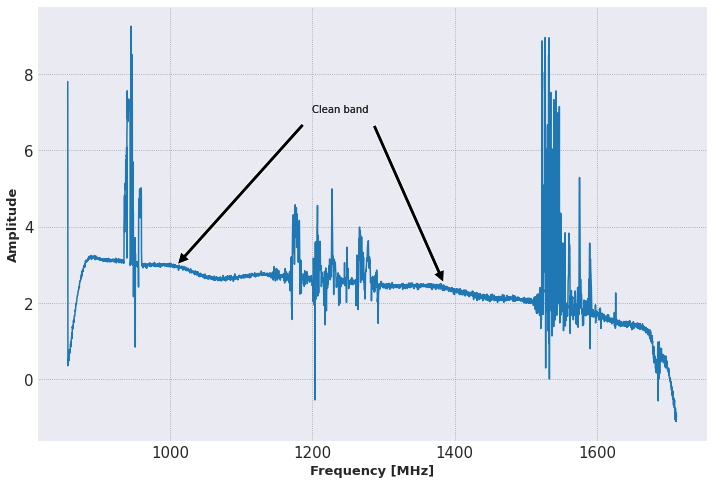}
  \caption{MeerKAT L-band bandpass. The black arrows are pointing to the less corrupt part of the band that we define as the ``clean band''. The ``clean band'' is between the the frequency band 970 - 1080 MHz and 1300 - 1500 MHz, excluding the GPS L3 band.}
  \label{vis}
\end{figure}

The data used for scientific analysis need to be from frequency bands that are less corrupted by RFI. In this section, we will, therefore, focus on analysing the RFI from the MeerKAT ``clean'' L-band wherein the spectrum is known to be less corrupted by RFI as depicted in Figure \ref{vis}. According to the Independent Communications Authority of South Africa (ICASA), the frequency band between 970 - 1080 MHz and 1300 - 1500 MHz are protected and reserved for radio astronomy purposes \footnote{\url{https://www.icasa.org.za/uploads/files/ITU-Reference-Review-the-Radio-Frequency-Band-Plan-31264.pdf}}. We therefore define those two regions as the ``clean'' band. However, the GPS L3 band (1375.9 - 1385.9 MHz) is confined within the upper ``clean'' band and was hence removed in our analysis.

\subsection{Clean L-band RFI Evolution}

\begin{figure}[htp]
  \centering
  \includegraphics[width=90mm]{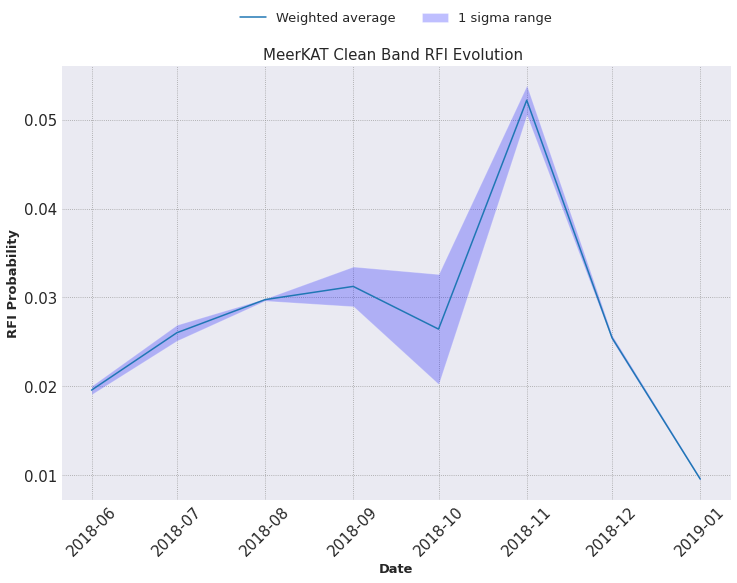}
  \caption{RFI occupancy as a function of month of the year 2018 and 2019 in the clean band. The blue line is the inverse-weighted-average (equation \ref{inverse_var_mean}) with the associated uncertainties depicted by the purple region. It can be noticed that the November RFI is 300$\%$ higher than that of the month of June.}
  \label{evolution}
\end{figure}

An increase in RFI occupancy level in the ``clean'' band poses a considerable challenge to radio astronomy because the data collected would be unusable . Therefore, it is essential to keep track of changes that might be happening over time in this ``clean'' band. Figure \ref{evolution} shows the change in fraction of data flagged over a period eight months. The dataset used consists of over 1500hrs of MeerKAT imaging data between 2018 and 2019.

To calculate the average RFI per month, we took into account the baseline length, as shown in  \cite{sihlangu}. Hence, for all months, we have used the same baseline lenghts. Consider a baseline index $j$, with $n$ being the number of observations. Then, the weighted average and the associated standard deviation can be calculated using equation  \ref{weighte_mean}  and \ref{standard_dev} respectively.

\begin{equation}
\theta_{j} = {\frac{\sum_{i = 1}^{n}t_{i,j}\times \theta_{j,k}}{{\sum_{i = 1}^{n}t_{i,j}}}}
\label{weighte_mean}
\end{equation}

\begin{equation}
\sigma_{j} = \sqrt{\frac{\sum_{i = 1}^{n}{t_{i,j}(\theta_{i,j} - \theta_j)^2}}{\frac{(M-1)}{M}\sum_{i = 1}^{n}t_{i,j}}}
\label{standard_dev}
\end{equation}

where $t_{i,j}$ is the observation length in seconds and  $\theta_{i,j}$ is the probability of RFI of a particular observation $i$ in a particular baseline $j$. The average was weighted using the length of the observation, such that long observation contribute more as compared to the short observations.

As a result, each month would have an array of RFI probability as a function of baseline length with the associated standard deviation. Finally, to get the average probability per month, we calculated the mean of the baseline array, taking into account the error associated with each $\theta_{j}$ by calculating the inverse-variance-weighted average and its associated variance using equation \ref{inverse_var_mean} and \ref{variance_inverse} respectively. The inverse-variance-weighting method is used when combining two or more values to minimise the weighted average variance, and each value is weighted in inverse proportion to its variance.

\begin{equation}
\theta_{Inv\_Var\_Mean} = \frac{\sum_{j} \frac{1}{\sigma^2_{j}}\theta_{j}}{\sum_{j}\frac{1}{\sigma^2_{j}}}
\label{inverse_var_mean}
\end{equation}

\begin{equation}
D^{2}(\theta_{j}) = \frac{1}{\sum_{j}\frac{1}{\sigma^2_{j}}}
\label{variance_inverse}
\end{equation}

We further calculate the weighted standard deviation as the square root of the weighted variance as shown in equation  \ref{error},

\begin{equation}
    \label{error}
    (S_\theta)_j =\sqrt{D^{2}(\theta_{j}) }
\end{equation}
\noindent

The blue line in  Figure \ref{evolution} is the inverse-weighted-average (equation \ref{inverse_var_mean}) with the associated errors depicted  by the purple region. It can be noticed that the November RFI is 300$\%$ higher than that of June. This is a significant increase in the clean band. As confirmed by the operation logs, the spike observed in November corresponded to when engineering activities were happening on-site.

Even though we cannot conclusively say that the engineering activities may cause the spike in November, there is a clear correlation between the engineering activities and the sudden rise of the amount of data flagged. 
Investigating the nature and locating the RFI activities is beyond the scope of this study. This will require diving into the MeerKAT's site activity logs and engineering job cards.  Furthermore, the MeerKAT data are archived on tape every 6 months and thus tedious to have recover these past data. 

To get a clear view of what was happening in November, we looked at a finer resolution (weekly probabilities). 
We found a considerable difference in RFI probabilities over the weeks of November, where we see up to 8$\%$ increase in some weeks, and in other weeks, the RFI occupancy is down to less than 2$\%$. The KATHPRFI framework provides us with the window to investigate and narrow down what could cause the change in RFI probabilities over time.

\begin{figure}[htp]
  \centering
  \includegraphics[width=90mm]{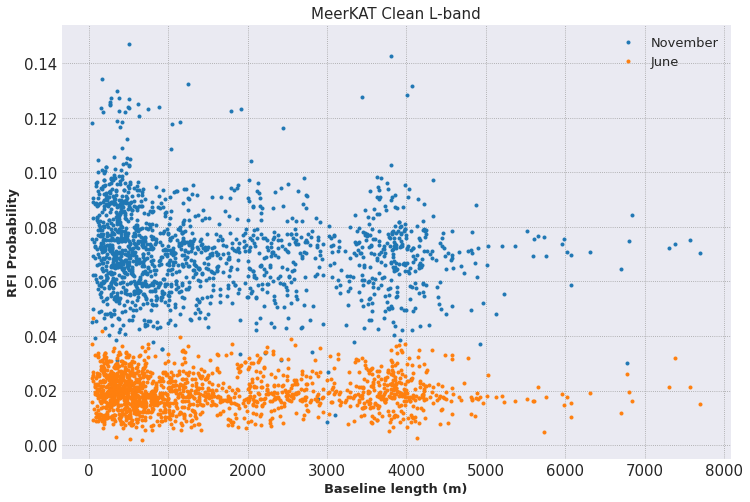}
  \caption{RFI occupancy as a function baseline length for the month of June and November.}
  \label{baseline_clean_band}
\end{figure}

We took a step further and investigated the distribution of RFI as a function of baseline length for June and November to check if we could see a difference between the two RFI distributions. Figure \ref{baseline_clean_band} shows the results, and we found that for the majority of the baselines, the probability in November is twice as much as in June. We also noticed that the likelihood of RFI for both months does not show an apparent reduction as the baseline length increases as one would expect. This may suggest that the RFI sources for the clean band are internal. Also, we noticed some few baselines where the RFI probabilities are considerably low for the month of November. We suspect that the observed jump in possibilities may result from a fault antenna during an imaging observation.

\subsection{DME RFI evolution during COVID-19 lockdown}

\begin{figure}[htp]
    \centering
    \includegraphics[width=90mm]{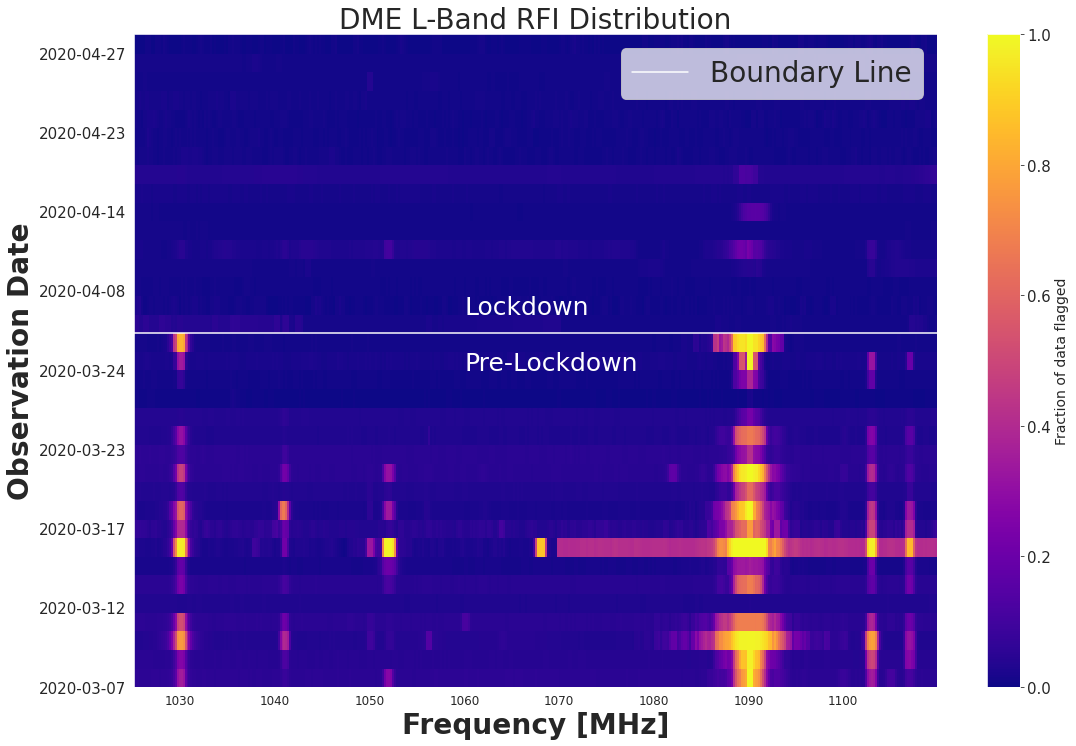}
    \caption{Fraction of the amount of data flagged before and during the lockdown as a function of DMEs frequencies. The white line shows a boundary between the pre-lockdown and the lockdown observations.}
    \label{dme}
\end{figure}

In \cite{sihlangu} it was argued that the increase of the RFI probability during the daytime is most probably due to the RFI coming from the flights' Distance Measuring Equipment (DMEs) as they pass across the site. During the hard lockdown due to the COVID-19 pandemic in early 2020, we carried out some observations to investigate the DME RFIs. The band 960-1215 MHz is reserved international for aeronautical radio-navigation service which is utilised for the operation and development of airborne communication with ground-based facilities. Figure \ref{dme} shows the fraction of data flagged pre-lockdown (08 March - 22 March 2020) through to the hard lockdown (08 April - 25 April 2020) as a function of frequency. The two main frequencies; 1030 MHz and 1090 MHz are used for ground-to-air and air-to-ground transmission respectively. One can notice a reduction and the disappearance of the DMEs RFI from those two frequencies during the hard lockdown. Further, we do see a clear reduction of all other frequencies that are confined within the band.

In order to check the impact of the DME RFI on fraction of data flagged, we computed the probability of RFI as a function of time of the day using lockdown and pre-lockdown observations. We found that on average more data was flagged over the the pre-lockdown (26\%) as compared to the lockdown period (21\%). This may suggest that the absence of the DME RFI significantly affect the amount of data flagged, hence, over the lockdown period less data was flagged as function of the time of the day and this findings are consistent to what has been shown and argued in \cite{sihlangu}.

\section{Conclusion}
Radio frequency interference mitigation and site characterisation is a challenge faced by radio astronomers. The RFI environment is worsening, and as more sensitive telescopes come online, the chance to pick faint RFI increases. While most researchers are focused on detecting and flagging RFI, we are  interested in cataloguing the nature and evolution of these outliers. The KATHPRFI framework has allowed us to dissect the MeerKAT RFI environment in both the UHF and L-band. The analysis we carried out allowed us to track the evolution of RFI as a function of time. 

We found that the RFI occupancy has increased significantly in November 2018 relative to other months. It remains unclear to us what has caused such a considerable drift in RFI occupancy in November but it correlates to the time where engineering activities where happening on site. However, we also found that there was a drastic decrease in DME RFI during the hard lockdown in South Africa due to the COVID-19 pandemic. 

The KATHPRFI can also be used to provide alerts about such sudden changes with this historical baseline known. This could be due to new sources of RFI or stem from any outliers in the data. These outliers could indicate telescope or correlator issues, just to mention a few. The KATHPRFI framework hence provides a window into the operational health of the telescope.

\section{Acknowledgements}
The MeerKAT telescope is operated by the South African Radio Astronomy Observatory, which is a facility of the National Research Foundation, an agency of the Department of Science and Innovation. This research has been conducted using resources provided by the United Kingdom Science and Technology Facilities Council (UK STFC) through the Newton Fund and the South African Radio Astronomy Observatory.

\end{document}